\documentclass[manuscript,screen,sigconf]{acmart}

\AtBeginDocument{%
  \providecommand\BibTeX{{%
    \normalfont B\kern-0.5em{\scshape i\kern-0.25em b}\kern-0.8em\TeX}}}

\setcopyright{rightsretained}
\copyrightyear{2022}
\acmYear{2022}
\acmDOI{10.5753/ise.2022.227048}

\acmConference[ISE '22]{ISE '22: The Second Brazilian Workshop on Intelligent Software Engineering}{October 04, 2022}{Uberlândia, MG}
\acmBooktitle{ISE '22: The Second Brazilian Workshop on Intelligent Software Engineering, October 04, 2022, Uberlândia, MG}

\usepackage{xspace}
\usepackage{tikz}
\usetikzlibrary{shapes.geometric,arrows,fit,matrix,positioning}

\newcommand{\SPIRA}[0]{SPIRA\xspace}

\begin{document}


\title[%
  SPIRA: Building an Intelligent System for Respiratory Insufficiency Detection%
]{%
  SPIRA: Building an Intelligent System for\break%
  Respiratory Insufficiency Detection%
}



\author{Renato Cordeiro Ferreira}
\orcid{0000-0001-7296-7091}
\affiliation{%
  \department{Institute of Mathematics and Statistics}
  \institution{University of São Paulo}
  \city{São Paulo}
  \country{Brazil}
}
\email{renatocf@ime.usp.br}


\author{Dayanne Gomes} 
\orcid{0000-0002-4982-7792}
\affiliation{%
  \department{Institute of Mathematics and Statistics}
  \institution{University of São Paulo}
  \city{Sao Paulo}
  \country{Brazil}
}
\email{dayanne@ime.usp.br}


\author{Vitor Tamae}
\orcid{0000-0001-8106-9576}
\affiliation{%
  \department{Institute of Mathematics and Statistics}
  \institution{University of São Paulo}
  \city{Sao Paulo}
  \country{Brazil}
}
\email{dai.tamae@usp.br}


\author{Francisco Wernke}
\orcid{0000-0002-6400-9202}
\affiliation{%
  \department{Institute of Mathematics and Statistics}
  \institution{University of São Paulo}
  \city{Sao Paulo}
  \country{Brazil}
}
\email{franwernke@usp.br}


\author{Alfredo Goldman}
\orcid{0000-0001-5746-4154}
\affiliation{%
  \department{Institute of Mathematics and Statistics}
  \institution{University of São Paulo}
  \city{Sao Paulo}
  \country{Brazil}
}
\email{gold@ime.usp.br}



\begin{abstract}
Respiratory insufficiency is a medic symptom in which a person gets a reduced
amount of oxygen in the blood. This paper reports the experience of building
\SPIRA: an intelligent system for detecting respiratory insufficiency from voice.
It compiles challenges faced in two succeeding implementations of the same
architecture, summarizing lessons learned on data collection, training, and
inference for future projects in similar systems.
\end{abstract}





\maketitle

\section{Introduction}\label{sec:introduction}

Respiratory Insufficiency is a medic symptom in which a person gets a reduced
amount of oxygen in the blood, which can lead to cough, tiredness, shortness
of breath, and in extreme cases, death. This symptom can be caused by many
diseases, such as the flu, severe asthma, or heart condition. In 2020, 
it became a sign of COVID-19 infection, before clinic tests were
available.

\break

\SPIRA is a project born during the \mbox{COVID-19} pandemic to use
Machine Learning (ML) to identify respiratory insufficiency via voice.
It involved medical doctors, linguistics, speech therapists, and computer
scientists. Modeling challenges were described in previous research%
~\cite{casanova-etal-2021-deep}, applying Deep Learning (DL) techniques.
This experience report focuses on challenges and lessons learned to build
an intelligent system that can be used in hospitals to pre-diagnose this
illness and collect data to enhance the model.

All the code as well as supplementary materials are open source and
freely available at: \url{https://github.com/spirabr}.

\section{The Architecture}\label{sec:architecture}

Intelligent systems are software systems that use artificial intelligence,
in particular machine learning, to achieve meaningful goals~\cite{Hulten2018}.
To manage the ML model lifecycle, intelligent systems are often divided into
two major workflows: training and inference.

\begin{figure}[hb]
\centering
\includegraphics[width=0.90\linewidth]{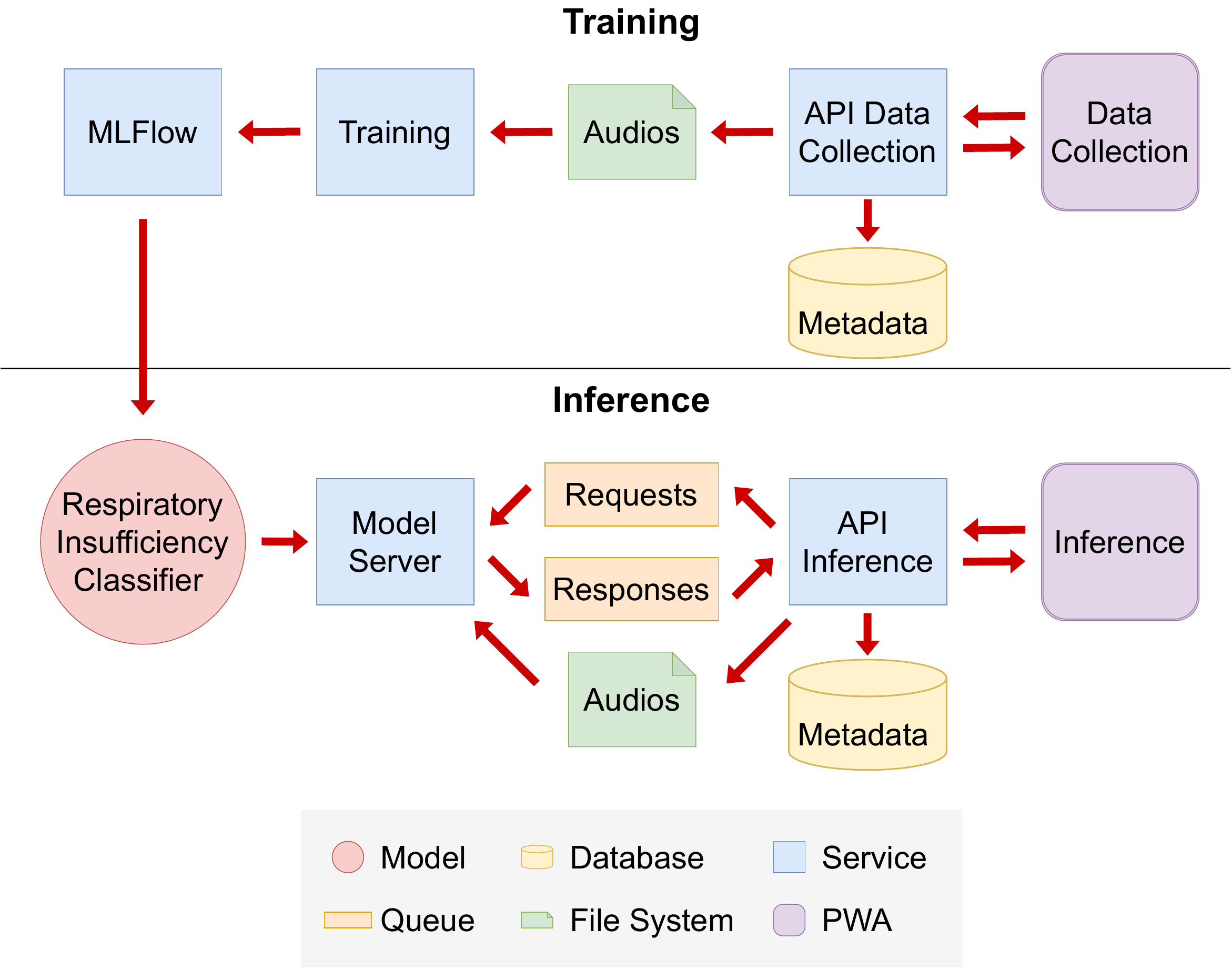}
\caption{The \SPIRA Architecture}
\Description{The \SPIRA architecture}
\label{fig:architecture}
\end{figure}

Given this requirement, \SPIRA was structured according to a microservices
architectural style~\cite{Boner:16}.
It was chosen for two reasons: system components can be implemented with
specialized software stacks, and developers can work in multiple components
at the same time. The \SPIRA architecture is illustrated in
\autoref{fig:architecture}.

Initially, collectors use the \textsc{Data Collection PWA} (Progressive Web App)
to gather ID, basic info and voice samples from volunteers, following a predefined
protocol. This data is sent to the \textsc{Data Collection API}, which stores
metadata in a document database and audios in a file system. Audios are
preprocessed to improve quality and reduce noise, then are finally
transformed into features. These features are used to train a
\textsc{Classifier} model, which is stored in the \textsc{MLFlow}
model repository. This workflow is experimental in nature and is executed by
a group of specialists to build better models from the available data.

Once the \textsc{Classifier} model is ready, it can be applied in unknown
data. The \textsc{Inference PWA} is connected with the \textsc{Inference API}
to collect data from potential patients. This time, each collection produces
an event in the \textsc{Requests} queue, which then is consumed by the
\textsc{Model Server}. This service applies the \textsc{Classifier} model.
The result is placed in the \textsc{Responses} queue, which is received
by the \textsc{Inference API} and shown by the \textsc{Inference PWA}.
Using a queue to manage requests and responses increases the system
resilience, since it ensures every collection will be processed eventually.

\section{First Version}\label{sec:1st_version}

This section describes the first version of \SPIRA, whose goal was to support
data collection to train the Deep Learning (DL) model described in previous
research~\cite{casanova-etal-2021-deep}.

\subsection{Data Collection}\label{subsec:1st_version_data_collection}


To reduce development cost, the initial proposal was to use a pre-existing
platform to collect data. \textit{WhatsApp}, a popular instant messaging app
in Brazil, was chosen for this task. Using this app offered several advantages:
  it had built-in audio recording and storage,
  it would save costs in deployment infrastructure, and
  it would require little to no effort to train data collectors
  (since most Brazilian know how to use the tool).
To avoid human error, collectors would use a chatbot to follow a protocol
predefined by specialists.


Unfortunately, preliminary tests showed that this solution would not meet
the project's audio quality requirement. \textit{WhatsApp} applies filters
that remove low-amplitude frequencies, and compresses files with lossy
methods using the Opus codec \cite{Hollerweger2012}. This reduces storage
and network consumption in mobile devices, but also erases signals useful
for training. Therefore, the chatbot was never developed.

This experiment showed that controlling the audio collection was critical
to create a training dataset. In the end, developers built a 
\href{https://doi.org/10.5281/zenodo.15824999}{website} in plain
HTML, CSS and JavaScript. Volunteers, mostly healthy, donated their audios,
but the tool could not be used by sick patients in hospitals. These experiences
influenced the second version of the data collection app, described in
\autoref{sec:2nd_version}.



\subsection{Training}\label{subsec:1st_version_training}

With a training dataset available, researchers produced their first deep
learning model~\cite{casanova-etal-2021-deep}. Since it was developed in
an experimental process, the training pipeline source code did not follow
any specific design patterns or conventions. Feasibility was the primary
goal rather than maintainability. Consequently, the model had no well-defined
interface: developers had to do many adaptations to create a single entry
point for predictions. In particular, input and output data processing
required most changes.



To deploy the model in an API for further research, developers had
to refactor the training pipeline. Nevertheless, this required extra
care to keep predictions coherent with the results presented in the
modeling paper~\cite{casanova-etal-2021-deep}. Therefore, this process
was lengthy, divided into three steps, to avoid introducing errors.

The first step was to check if all dependencies were compatible
with those from the article version. This was necessary because
many machine learning libraries are under heavy development and
can change their behavior between versions.
The second step was to create a set of automated tests to isolate
errors in the code. Data transformation functions underwent most changes,
so most new tests covered them to ensure their correctness.
Finally, the third step was to do a more detailed analysis of results.
Unfortunately, true reproducibility was impossible, since the researchers
did not use any experiment tracking tool while creating their preliminary
model. Therefore, there was no way to recover the exact trained model used
in the experiments. Consequently, the disparity in metrics was minimized,
but not eliminated.


After finishing all adaptations in the training code, the new model
was deployed via a \href{https://doi.org/10.5281/zenodo.15824999}{HTTP API},
supporting responding to inference requests and retrieving the inference history.
All results were stored in a local database, which was used for a more thorough analysis of the model.

The lack of an explicit interface for the model delayed considerably
the development of the server system. Furthermore, the server got
highly coupled with the model, since the input data processing had to
be reimplemented both in the training and server code. Finally,
the lack of automated tests, particularly for data processing
functions, made it difficult to debug errors and avoid regressions.
These experiences influenced the second version of the server system,
described in \autoref{sec:2nd_version}.



\section{Second Version}\label{sec:2nd_version}

The implementation described in the last section provided valuable lessons
to improve \SPIRA. This section describes the second version of the system,
which implements the intelligent system architecture presented in
\autoref{fig:architecture}. Its goal was to produce an improved model:
  trained with data collected from hospital patients, and
  integrated into an app for pre-diagnosis of respiratory insufficiency.

\subsection{Data Collection}\label{subsec:2nd_data_collection}

This section explains the context, challenges, lessons learned, and results
in the process of creating a new data collection app for \SPIRA. The source
code is available at: \url{https://github.com/spirabr}.



\subsubsection{Context}\label{subsubsec:2nd_data_collection_context}


The second version of \SPIRA aimed to collect data from healthy and sick
volunteers in hospitals. To accomplish this, the researchers created a new
protocol for data collection: for each participant, the app should register
some basic info -- such as the volunteer's hospital ID -- followed by multiple
audio recordings. Later on, the resulting dataset could be cross-referenced
with the hospital's medical records to label its entries.


The collection protocol was defined in a joint effort between phonoaudiology
and linguistics researchers. The goal was to capture biomarkers in the speech
indicating respiratory insufficiency. From a software standpoint, this meant
that audio signals should be captured raw, with no compression or filtering.

Since the data collection was going to happen in hospitals, the app had
to be used by known data collectors, such as nurses or graduate students.
To reduce costs, the researchers proposed using the collectors' own mobile
phones as devices for the collection, thus relying on the quality of their
microphones to capture audio.

Fortunately, using known data collectors allowed training them beforehand
to apply the collection protocol, making the app's UX simpler. Nevertheless,
\SPIRA's designer had to make applying the protocol as quick as possible,
lowering the chance volunteers would start and then give up of participating.
To accomplish this, the designer worked with the researchers to create a
wireframe%
\footnote{The wireframe was made with Adobe XD and is available at:
\url{https://web.archive.org/web/*/https://xd.adobe.com/view/0994d1dd-06bf-4e0b-ae5e-254abb08e5af-5cc7/}}
showing all screens that should be included in the app.

As described in \autoref{sec:1st_version}, using an existing platform such
as \textit{WhatsApp} did not provide much control over audio recording. Therefore,
the developers chose to build a \href{https://doi.org/10.5281/zenodo.15825123}{%
Progressive Web App (PWA)} \cite{RichardLePage2020}, which allowed them to use
Web development technologies while maintaining more control over the phone's
operating system. The development stack was composed by:
\begin{enumerate}
  \item An instance of \href{https://mongodb.com}{MongoDB}, a NoSQL document
        database to store metadata about volunteer participants. This type of
        database provided a flexible schema that could be evolved during the
        implementation of the collection protocol.

  \item A \href{https://doi.org/10.5281/zenodo.15825207}{back-end service}
        developed with \href{https://quarkus.io}{Quarkus},
        a Java Framework built for containerization. This framework provided
        all resources to build a web server, while also making it easier to
        deploy them in any environment.

  \item A \href{https://doi.org/10.5281/zenodo.15825123}{front-end service}
        developed with \href{https://vuejs.org}{Vue.js},
        a JavaScript framework focused on creating iterative web pages.
        This framework allowed creating an installable Progressive Web App
        (PWA) effortlessly.
\end{enumerate}

\subsubsection{Challenges}\label{subsubsec:2nd_data_collection_challenges}

There were three main challenges in building the data collection app:
the UX design, the audio recording, and the system deployment.

Since volunteer participants were in a stressful situation in hospitals,
the data collection had to prioritize quickness. The designer translated
the collection protocol in an intuitive UI, keeping only the essential steps.
Screen scrolling was avoided at all costs, as it could lead to a slower
experience and could even interfere in a part of the protocol that required
reading. All data collected was saved locally, avoiding any loading time
due to slow connections (to be expected inside hospitals). Only when the
collector had internet access, then the app would make requests to sync
its cache.

At first sight, recording audio in a web app can seem straightforward.
In practice, many decisions have to be taken while recording audio.
The browsers' default recording API, for example, does not support
WAV encoding by default~\cite{AudioRecorderRFC}. As a result, developers
need to find a library that can record WAV files that record Android's
Pulse Code Modulation (PCM) string of bytes \cite{AndroidRecordingFormatAPI}.
In addition, web environments usually assume resource scarcity and
try to compress any large files. This way, the default web recording
API applies a standard codec (Opus codec) to send files to servers.
While building the data collection app, the developers found a library
worked around both problems:
\href{https://www.npmjs.com/package/extendable-media-recorder}{extendable-media-recorder},
a NodeJS module that extends the standard audio recording API.

While delivering the system, the back-end service was deployed on-premise
and the front-end in a CDN (Content Delivery Network). The on-premise server
and the cloud PWA were accessed by the same URL. A \href{https://nginx.com/}{Nginx}
proxy server routed each request. The main challenge was to keep the server stable
to receive all requests. Often times, a stable connection to the back-end was
not possible. Thanks to the PWA capabilities, it was easier to provide an
offline version of the app, storing all recordings locally. Moreover,
the app featured a page where the collector could see all audios that
were not sent yet (and which would be sent only if they were online).
This improved the app's resiliency, avoided losing data, and improved the UX.

\subsubsection{Lessons Learned}\label{subsubsec:2nd_data_collection_lessons_learned}

This section summarizes lessons learned while building the data collection
app, regarding the use of PWAs and audio recording.

Using Progressive Web Apps was very practical since most developers nowadays
learn web development. It provided the experience of a native app without the
burden of learning mobile-specific languages and tools. Notwithstanding, PWAs
did not provide fine-grained control over the hardware, specifically for audio
recording. Overall, PWA features allowed an easy-to-make offline experience,
storage efficiency, and a UX boost by installing the app in the phone.

The library to record WAV files in web environments was convenient and solved
many challenges related to recording. It was not necessary to do any adaptations
to use the library, since it extends the default audio recording API -- only
some configurations were needed. One possible drawback was that volume
correction options are disabled by default, which occasionally makes
the audio less audible for human ears. Nevertheless, since the goal
was to find biomarkers in the speech, capturing the audio raw was
preferable for this application.

\subsubsection{Results}\label{subsubsec:2nd_data_collection_results}

By the publication of this paper, the data collection app has been
used to make 2125 collections in five partner Brazilian hospitals.
This data will be cross-referenced with the patient's medical record.
The resulting dataset will allow a new experimental phase, whose goal
is to further improve the respiratory insufficiency detection model.

\subsection{Inference}\label{subsec:2nd_inference}

This section explains the context, challenges, and lessons learned in the
process of creating a new inference system for \SPIRA. The source code is
available at: \url{https://github.com/spirabr}.

\subsubsection{Context}\label{subsubsec:2nd_inference_context}


The second version of the model, using multiple voice recordings, is already
under development. This new experimental phase will provide multiple candidate
models. The goal is to test them in hospital environments to help pre-diagnose
patients with respiratory insufficiency. To achieve this goal, it is necessary
to create a new system for testing and versioning multiple models.

As described in \autoref{sec:1st_version}, the lack of a contract between
the model and the server makes it impractical to generalize one server
implementation for multiple models. Therefore, defining a clear interface
is essential for the second version. Furthermore, adding a model registry
can provide many advantages regarding reproducibility and testing.

\subsubsection{Challenges}\label{subsubsec:2nd_inference_challenges}

There were four main challenges in building the inference system:
response resiliency, service architecture, coupling between API and server,
and reproducibility.

Once predictions are requested to the inference API, they have to be
processed by the models, regardless of the conditions of the server
or the time needed to process the request. The solution for this was
to use an \href{https://doi.org/10.5281/zenodo.15825105}{asynchronous
event-driven API}, since requests can be stored in a message broker and
wait until the \href{https://doi.org/10.5281/zenodo.15825111}{server} is
ready to process them.
The \href{https://doi.org/10.5281/zenodo.15825144}{message service} solution
chosen was NATS, due to its scalability with the use of clusters and its
self-healing features, which provide a higher resilience to the system.

The MVC architectural pattern was initially adopted for the API service.
However, the need for an event-driven API challenges this approach.
The \href{https://doi.org/10.5281/zenodo.15825105}{inference API}
has two entry points: it responds to client requests via a HTTP API
built with \href{https://fastapi.tiangolo.com}{FastAPI},
and it updates a \href{https://mongodb.com}{MongoDB} document database
with data from messages received via the \href{https://nats.io}{NATS}
message broker.

This requirement led the developers to choose another architectural
pattern: the hexagonal architecture~\cite{cockburn}.
This pattern divides the application into three parts:
  the core, containing all the business logic of the application;
  the adapters, encapsulating all external dependencies required by the core; and
  the ports, defining the interface between adapters and core.
This way, all business logic can be centralized in the core regardless
of the origin of the request (via events or via the HTTP protocol),
whereas the MVC implementations usually focus only on the HTTP API.

As discussed in \autoref{sec:1st_version}, a
\href{https://doi.org/10.5281/zenodo.15825160}{well-defined interface}
is necessary to avoid coupling between the server and the model.
This way, the server can be used with any other model following the same
prediction interface. By applying the hexagonal architecture in the server,
this interface can be easily abstracted as a port. It becomes the contract
to be used between a prediction use case inside the core and the real model
stored in the model registry adapter.

Improving reproducibility was also a major concern discussed in
\autoref{sec:1st_version}. To systematically manage the life-cycle
of machine learning models, it is necessary to track experiments,
store parameters, and version the training pipeline source code.
To achieve that, the solution was to adopt \href{https://mlflow.org/}{MLFlow},
a model registry platform that is compatible with various machine
learning libraries.

\subsubsection{Lessons Learned}\label{subsubsec:2nd_inference_lessons_learned}

This section summarizes lessons learned in the ongoing development of
the inference system, regarding the use of message brokers and hexagonal
architecture.

The NATS server became a great addition to the system, solving all the
requirements for a message service solution. The choice of a message broker
over a queue system was justified because the server system can request
predictions for multiple models. With a broker, there is no need to
create additional infrastructure besides new model servers, whereas
with queue systems, each model service would require its own queue.

The hexagonal architecture addressed well the requirement of
using both events and HTTP requests in the services. Moreover, an
\href{https://doi.org/10.5281/zenodo.15825160}{interface for the
prediction model} emerged naturally in this architectural pattern
(as a port). On that note, it is important to decide beforehand if the
conversion between core and adapters will be delegated to the ports,
i.e., whether the ports will be interfaces implemented by the adapters
(object-oriented style) or standalone classes and functions (functional style).
Besides personal preference, the latter makes it easier to isolate unit tests
for these conversions, whereas the former can be less verbose.

\section{Conclusion}\label{sec:conclusion}

Respiratory Insufficiency is a medic symptom in which a person gets a reduced
amount of oxygen in the blood. This symptom can be caused by many diseases,
such as the flu, severe asthma, or heart condition. This paper reported
the experience of building \SPIRA: a multidisciplinary project created
in the \mbox{COVID-19} pandemic that uses Deep Learning techniques to
identify respiratory insufficiency via voice~\cite{casanova-etal-2021-deep}.

The paper presented the architecture used as a guide to build an intelligent
system that could be used as a pre-diagnosis tool inside hospitals.
It described two succeeding implementations of the architecture,
highlighting challenges and lessons learned from each step.
The challenges ranged from UX design, audio recording, and system
deployment -- for data collection -- to response resiliency,
service architecture, coupling, and reproducibility -- for inference.
Overcoming them led to many lessons learned regarding the use of PWAs,
audio recording, message brokers, and hexagonal architecture. These
experiences can be useful for researchers and practitioners who may
work with similar data and conditions.

The data collection app is currently in production, and it is being used
to create the new dataset. By mid-2022, the second experimentation phase,
led by \SPIRA researchers, is about to start. The first version of the
inference system is expected to be deployed in late 2022. Given that
data collection and inference will require the same data collection protocol,
the existing data collection PWA can be generalized for use in the inference
process. When deployed, the inference PWA will be used to test multiple
models in hospitals. When the best model gets chosen, the second version
of the system will be done. It then can be made available to help
pre-diagnose respiratory insufficiency in hospitals.



\begin{acks}
This work was supported by FAPESP project 2020/06443-5 (SPIRA). The authors
also would like to thank the developer team who volunteered their work during
the COVID-19 pandemic.
\end{acks}



\bibliographystyle{ACM-Reference-Format}
\bibliography{main}


\appendix

\end{document}